\documentclass[pra,twocolumn,showpacs,superscriptaddress]{revtex4}

\usepackage{amsmath,amssymb}
\usepackage{graphicx}
\usepackage{mymacros}

\begin{document}

\title{Steering quantum information in driven qubit chains}
\title{Quantum router based on ac control of qubit chains}

\author{David Zueco}
\email{David.Zueco@physik.uni-augsburg.de}
\affiliation{Institut f\"ur Physik, Universit\"at Augsburg,
        Universit\"atsstra{\ss}e~1, D-86135 Augsburg, Germany}
\author{Fernando Galve}
\email{Fernando.Galve@physik.uni-augsburg.de}
\affiliation{Institut f\"ur Physik, Universit\"at Augsburg,
        Universit\"atsstra{\ss}e~1, D-86135 Augsburg, Germany}
\author{Sigmund Kohler}
\affiliation{Instituto de Ciencia de Materiales de Madrid, CSIC,
	Cantoblanco, E-29049 Madrid, Spain}
\author{Peter H\"anggi}
\affiliation{Institut f\"ur Physik, Universit\"at Augsburg,
        Universit\"atsstra{\ss}e~1, D-86135 Augsburg, Germany}

\date{\today}

\begin{abstract}
We study the routing of quantum information in qubit
chains. This task is achieved by suitably chosen time-dependent local
fields acting on the qubits.  Employing the physics of coherent
destruction of tunneling, we demonstrate that a
driving-induced renormalization of the coupling between neighboring
qubits provides the key for controlling the transduction of
quantum information between permanently coupled qubits.
We employ this idea for building a quantum router.
Moreover,  we discuss the experimental implementation with Penning
traps, and study the robustness of our protocol under realistic
experimental conditions, such as fabrication uncertainties and
decoherence.
\end{abstract}

\maketitle

\section{Introduction}

Entanglement, the inextricable quantum correlation between
two physical objects, is a key resource for quantum informational
tasks which promise pushing further our current computational limits
\cite{Deutsch1985}. This resource, which is produced through the
interaction of two (or more) quantum objects, needs to be created,
used and transferred under strict control such that a successful
computation can be performed. Transporting entanglement may be
considered as the analogue of electric currents for classical
computation, which need to be produced, then used for computation
inside a CPU, and finally transferred to the output device.

In this paper we will focus on the process of transferring information
once the computation has been performed. While long distance quantum
communication can be done quite efficiently with photons
\cite{Zeillinger2007}, communication inside a quantum computer will
probably use linearly arranged solid-state qubits \cite{bose03,
Bose2006}. Lately, much attention has been paid to spin chains with
nearest neighbor coupling, which allow for perfect transfer of
quantum information \cite{Christandl2004, Nikolopoulos2004,
diFranco2008, wu2009}. The simplicity of such chains and the fact that
they do not require special control make them robust for experimental
implementations \cite{diFranco2008b}. It is our  belief that a quantum
computer will profit from more complicated devices which allow not
only linear propagation, but also the possibility to stop the signal
at a given node in a network, reverse its motion, etc., such that the
information can be \textit{routed} to the desired node with minimal
control. This could be particularly useful in quantum registers to
store, retrieve, and transfer information contained in different
qubits.

The possibility of controlling the transfer in a spin chain is
exciting.  Such a chain could be used to inhibit access to parts of the
register, or even to divide a signal into two different branches for
distribution in a network of nodes, provided that a particular
propagation direction can be chosen.  At this point one could argue
that for forbidding the propagation through a particular direction we
``just need'' to switch off the coupling in that direction
\cite{romeroisart2007}.  However, coupling and decoupling qubits is a
formidable task  \cite{You2005, Niskanen2007, zagoskin2007,
Mariantoni2008a}.  Therefore we should think of an alternative for
breaking the symmetry and, thus, allowing the transfer in one
direction, without modifying the inter-qubit couplings.

The mathematics relating to coherent destruction of tunneling
\cite{Grossmann1991, Grossmann1992, GrifoniPR} or dynamical
localization, i.e., the suppression of ballistic transport by ac
fields \cite{GrifoniPR, kenkre2000, PlateroPR, Kohler2005a,
Longhi2008} suitably solves the objective in the high-frequency limit
\cite{Kayanuma2008}. Such driving renormalizes the tunnel matrix
elements and has been investigated previously in the context of
ac-driven mesoscopic transport \cite{Camalet2003a, PlateroPR,
Kohler2005a}.  In a recent theoretical work \cite{Creffield2007}, it
has been suggested to temporarily suppress tunneling between specific
neighboring sites of an optical lattice, such that a particle will be
transported into a direction of choice.
Here we will demonstrate how to use such a dynamical renormalization
for transmitting quantum information.

In Sec.~\ref{sec:router} we introduce the concept of a \textit{quantum
router} and present in Sec.~\ref{sec:cdt} an according chain
model together with brief explanation of the coherent destruction of
tunneling on which our quantum router relies.  Section
\ref{sec:protocol} is devoted to the development of an according
protocol for entanglement distribution and state transfer.  A proposal
for the implementation in Penning traps together with an analysis of
the robustness of the method under experimental conditions is
presented in Sec.~\ref{sec:implementation}.

\section{A router for quantum information}
\label{sec:router}

Quantum information processing protocols usually depend on having full
control of the qubits and their mutual interactions.  This control
relies on external devices such a lasers or signal generators, which
limits the number of qubits by space considerations.  One
possible solution of this problem is splitting the processor into
small blocks which, however, need to be connected
\cite{Burgarth2006}.  Increasing the complexity of the network
requires the selective coupling and decoupling of the blocks. We
term such a coupler/decoupler \textit{quantum router}.

\begin{figure}[tb]
 \includegraphics[width=6.2cm]{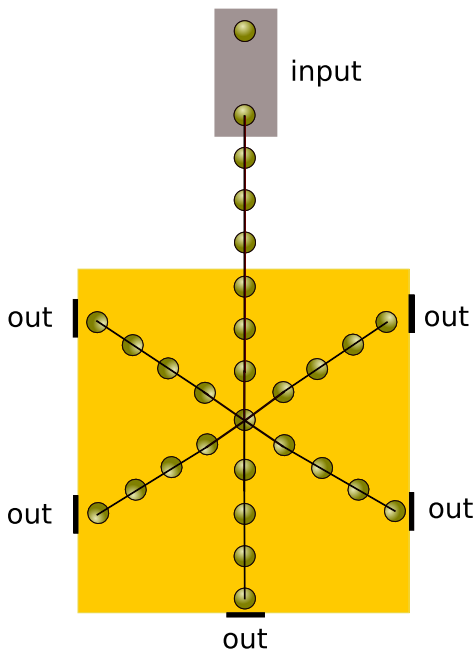}
\caption{(Color online) Hub configuration. One qubit of an entangled pair is
connected to a line with perfect transfer. Then the entanglement propagates
and is later steered to one of the possible outputs, i.e.,
distributed to the desired transmission line. This enables
quantum communication inside the quantum computer by both direct
state transfer and by teleportation once the entanglement of distant
parties has been established.}
\label{fig0}
\end{figure}%
Let us assume for the moment that we have complete control of the
coupling between every pair of qubits in an arbitrary array, despite
that fact that this requires quite sophisticated engineering, at least
much more than the perfect, unidirectional entanglement transfer
described in Refs.~\cite{Christandl2004, Nikolopoulos2004,
diFranco2008}. Thus such a specialized piece of architecture is
probably useful only for very specific tasks at particular points
during computation, transfer, and storage.  One example is the hub
sketched in Fig.~\ref{fig0}, let it be realized by a passive or an
active element.  When entanglement arrives at the junction, it should
be routed to an output of choice. The output channel may either be
selected by the router itself or depend on an additional classical signal containing that information.  Moreover, the
router may select the output direction by the time at which the signal
arrives at the hub.  This means that the hub effectively opens the
channels in turns and delivers the signal accordingly.

\begin{figure}[tb]
 \includegraphics[angle=90,width=6cm]{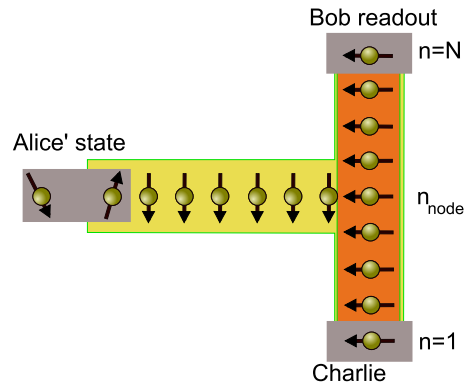}
\caption{(Color online) T-shaped configuration. Alice sends the state of one
of her qubits without dispersion along the horizontal chain.
When the state arrives at the node, an ac field with a
ratchet-like profile transfers the signal to Bob or to Charlie.}
\label{fig:T}
\end{figure}

A main obstacle for the experimental realization of a router is that
it might be difficult if not impossible to simply turn off the
interaction between neighboring qubits.  Still, there exists an
indirect way for this task that is based on the physics of ``coherent
destruction of tunneling''.  Let us consider a chain of qubits with
given qubit-qubit interaction determined by the experimental
setup.  But nevertheless there exists the possibility of controlling
the energy splitting of each qubit, which is typically less demanding
than controlling the coupling.  Then, as we will demonstrate below, an
ac field with proper amplitude and frequency acting on the qubit will
effectively renormalize the qubit-qubit interaction.  It is even
possible to suppress this interaction almost entirely.  After explaining
the underlying mechanism, we apply the idea to the T-shaped setup
sketched in Fig.~\ref{fig:T}, where Alice sends a quantum state
through the horizontal quantum channel.  When the state arrives at the
node, it shall be directed either to Bob or to Charlie.  We
demonstrate below that this can be achieved by applying to the qubit
splittings of the vertical chain an ac control with a ratchet-like
amplitude distribution.

\section{Coherent destruction of tunneling in chains of qubits}
\label{sec:cdt}

We consider entanglement transfer in the vertical qubit chain of
Fig.~\ref{fig:T} modelled by the Hamiltonian
\begin{equation}
\label{iso-gral}
 H = \frac{1}{2}\sum_n h_n(t) \sigma_n^z
    + \frac{J}{4}\sum_n \left(\sigma_n^+\sigma_{n+1}^-
      +\sigma_n^-\sigma_{n+1}^+\right)
\end{equation}
with $\sigma^{\pm}_n = \frac{1}{2} (\sigma^x_n \pm \iu \sigma^y_n)$
and $\sigma^\alpha_n$, $\alpha =x,y,z$ the usual Pauli matrices for
spin $n$, while $\hbar=1$.  This Hamiltonian is known as the isotropic XY
model with time dependent local fields $h_n(t)$. The qubit-qubit
coupling is a standard interaction, realized either by direct
engineering of the chain or as a rotating-wave approximation (RWA)
where terms of the type $\sigma_n^+\sigma_{n+1}^+$ and
$\sigma_n^-\sigma_{n+1}^-$ are neglected
\cite{Ciaramicoli2003,Stahl2005, Helmer2007a}. By means of the
Wigner-Jordan transformation, Hamiltonian \eqref{iso-gral} is
equivalent to the one-dimensional spinless fermion model described by
the Hamiltonian
\begin{equation}
H=\frac{1}{2}\sum_n h_n(t) c_n^\dagger c_n + \frac{J}{4}\sum_n \left(c_n^\dagger c_{n+1}+c_{n+1}^\dagger c_n\right).
\label{excit}
\end{equation}
Furthermore, we will see that the mechanism explained below can be
observed as well in a bosonic chain of harmonic oscillators for which
the Hamiltonian reads
\begin{equation}
H = \frac{1}{2}\sum_n \frac{p_n^2}{m} + m\omega_n^2(t)x_n^2
    + \lambda\sum_n x_n x_{n+1},
\label{osci}
\end{equation}
provided a RWA in the coupling between the oscillators can be applied,
i.e., neglecting the counter-rotating terms $a_n a_{n+1}$ and
$a_n^\dagger a_{n+1}^\dagger$.

We assume that the onsite qubit splitting possesses a time dependence
stemming from a classical field of the form
\begin{equation}
 h_n(t)=A b_n\cos(\omega t) \; .
\end{equation}
In the interaction picture defined by $\widetilde X = U_0^\dagger(t) X
U_0(t)$, where $U_0(t) = \exp[-i\varphi(t)\sigma_z]$ and
$\dot\varphi(t) = h_n(t)/2$, the Hamiltonian (\ref{iso-gral}) reads
\begin{equation}
\widetilde H(t)
=
J \sum \e^{\iu \eta_n (t)} \sigma_n^+\sigma_{n+1}^-
+ \e^{-\iu \eta_n (t)} \sigma_n^-\sigma_{n+1}^+
\label{tildeH}
\end{equation}
with the phase
\begin{equation}
\eta_n (t) =  \frac{A}{\omega}  (b_n - b_{n+1}) \sin(\omega t) .
\end{equation}
In the high-frequency limit $\omega \gg J$, the interaction-picture
Hamiltonian \eqref{tildeH} can be replaced by its time-average.
Thus, we obtain an
effective static XY chain Hamiltonian for which the coupling between
qubits $n$ and $n+1$ is renormalized according to
\begin{equation}
\label{Jeff}
 J \to J_{\text{eff},n}
= J J_0 \left ( A \frac{b_n-b_{n+1}}{\omega} \right ),
\end{equation}
where $J_0$ denotes the zeroth-order Bessel function of the first
kind.  Thus, the coupling between sites $n$ and $n+1$
can be tuned to zero by choosing
driving parameters for which $A(b_n-b_{n+1})/\omega$ is a root of
$J_0$.  This renormalization has been used to explain the
phenomenon of coherent destruction of tunneling \cite{Grossmann1991,
Grossmann1992, Longhi2007, GrifoniPR}.  Here we use it to steer
quantum information by temporarily suppressing the interaction between
two particular spins.

\section{Protocols and control mechanism}
\label{sec:protocol}

Quantum routers may be useful for various purposes.  We focus in the
following on two of them, namely entanglement distribution and quantum
state transfer.  We develop for these aims protocols that rely on the
CDT mechanism introduced in the previous section.  It will turn out
that both protocols are closely related.

\subsection{Entanglement routing}
\label{sec:routing}

An essential resource for quantum communication is a pair of distant entangled
qubits, one at each end of the communication channel \cite{teleport}.  Such
pairs can be obtained by creating the entangled pair locally and
subsequently transfer to each end one partner.  Recently, Creffield
suggested to use the CDT mechanism described above for the controlled
coherent propagation in a lattice \cite{Creffield2007}. Different to this
entanglement transfer is the previously suggested non-local creation
of entanglement in a spin chain \cite{Galve2009a, bose}.

The topic of this subsection, by contrast, is to create an entangled
pair at Alice's place and to route one partner to either Bob or
Charlie, see Fig.~\ref{fig:T}.  A possible protocol for this task is
the following: Let Alice have a two-qubit entangled state
$\ket{\psi}_\mathrm{A}$ that she wants to share with Bob.  She will then
attach one of the qubits to a chain with isotropic nearest neighbor
interaction such that the qubit propagates towards the node with the
vertical chain.  If she for example owns the state $\ket{\psi(0)}
= (\ket{01}+\ket{10})/\sqrt{2}$, the horizontal chain has
the initial state
\begin{equation}
\ket{\psi}_\mathrm{in}
= \frac{1}{\sqrt{2}} \big ( \ket {1}_{A} \ket{00\ldots00}
                           +\ket {0}_{A} \ket{10\ldots00}
                     \big ) ,
\end{equation}
while the desired final state is
\begin{equation}
\ket{\psi}_\mathrm{out}
=\frac{1}{\sqrt{2}}\big( \ket {1}_{A} \ket{00\ldots00}
                        +\ket {0}_{A} \ket{00\ldots01}
                   \big) .
\end{equation}
Such perfect entanglement transfer can be achieved by specific static
couplings \cite{Christandl2004, Nikolopoulos2004, diFranco2008}.  The
question is now whether time-dependent fields allow one to route the
qubit in a controlled manner from the node of the T-junction to either
Bob or Charlie.

When the travelling qubit arrives at the node of the T-junction (see
Fig.~\ref{fig:T}), coherent destruction of tunneling must become
active such that propagation to, say, Charlie's branch is suppressed.
Our goal is now to find an ac field that such that the entanglement is
perfectly transmitted to Bob.  Moreover, the protocol should be flexible
enough to allow routing to Charlie as well.

\begin{figure}[tb]
\includegraphics[width=6.5cm]{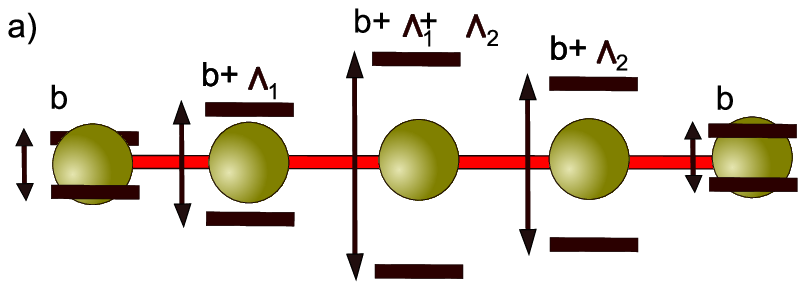}

\vspace{3ex}
\includegraphics[angle=-0,width=7.cm]{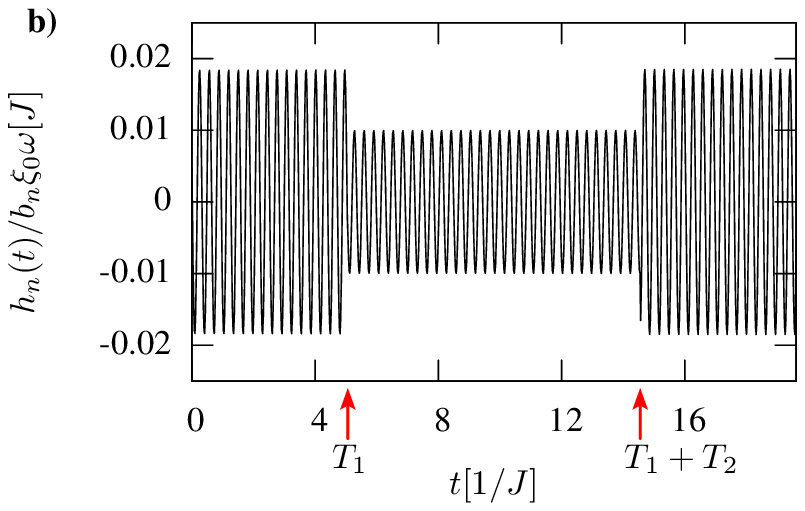}
\caption{(a) Spatial variation of the local qubit energy splittings
along the vertical chain.  It is chosen such that the absolute value
of differences between two neighboring sites alternates between
$\Lambda_1$ and $\Lambda_2$.
(b) ac field with alternating driving amplitude.  The amplitudes
[see Eq.~\eqref{coseno}] are such that tunneling between qubits with
either energy difference $\Lambda_1$ or $\Lambda_2$ is suppressed.
}
\label{driving}
\end{figure}

We consider now a qubit chain that consists of blocks with four qubits
with the ratchet-like energy splittings \cite{Astumian2002, RMP2009}
\begin{equation}
 b_{n} =
\begin{cases}
b,   & n=4n'
\\
b + \Lambda_1,   & n=4n'+1
\\
b + \Lambda_1 + \Lambda_2,   & n=4n'+2
\\
b + \Lambda_2,   & n=4n'+3
\end{cases}
\label{campos}
\end{equation}
for integer $n'$, as is sketched in Fig.~\ref{driving}(a).  It is
constructed such that the energy differences between two neighboring
qubits are given by the sequence
$\Lambda_1,\Lambda_2,-\Lambda_1,-\Lambda_2$, i.e., their absolute
values alternate between $\Lambda_1$ and $\Lambda_2$.  CDT can now be
employed for temporarily suppressing tunneling between qubits with the
one or the other value of the splitting [note that $J_0(-x)=J_0(x)$,
such that the CDT condition is not sensitive to the sign of
$\Lambda_{1,2}$].  A field that alternates between these two
possibilities can induce directed transport \cite{Creffield2007}.  We
here use this idea to route entanglement.  To be specific, we act on
qubit $n$ the driving field
\begin{equation}
 h_n(t)=
\xi_0 \omega b_n \cos (\omega t) \times
\begin{cases}
  1/\Lambda_1, &  0\leq t < T_1
  \\
  1/\Lambda_2, &  T_1 \leq t < T_1+T_2
\end{cases}
\label{coseno}
\end{equation}
and periodically continued, see Fig.~\ref{driving}(b), where
$\xi_0=2.4048\ldots$ is the smallest positive root of the Bessel
function $J_0$.  For this field, the CDT condition $J_\text{eff}=0$ is
fulfilled the transitions $4n' \leftrightarrow 4n'+1$ and
$4n'+2\leftrightarrow 4n'+3$ during the time interval $[0,T_1]$, while
the other transitions are still allowed.  For them the effective
coupling energy is $J_{\text{eff},1} = J
J_0(\xi_0\Lambda_{1}/\Lambda_{2})$.  The length of the time interval,
$T_1 = \pi/|J_{\text{eff},1}|$, corresponds to a half tunnel cycle
between qubits $4n'$ and $4n'+1$, such that these qubits exchange their
state.  During time interval $[T_1,T_1+T_2]$, we obtain that the
coupling between spins with energy difference $\Lambda_2$ is
suppressed, while the other coupling is renormalized according to
$J_{\text{eff},2} = J J_0(\xi_0\Lambda_{2}/\Lambda_{1})$.  Thus
the second stage has to last for $T_2 = \pi/|J_{\text{eff},2}|$.

Next, we test numerically the quality of the entanglement transfer.
Since the isotropic XY model conserves the number of excited spins, it
is sufficient to consider only the subspace with at most one qubit in
state 1, while all other qubits are in state 0.  Then the chain can be
mapped to the tight-binding model
\begin{equation}
\label{eq1}
H = \sum_n h_n(t) |n\rangle\langle n|
   + \frac{J}{2}\sum_n \left(|n\rangle\langle n+1|+|n+1\rangle\langle n|\right),
\end{equation}
with the $(N+1)$-dimensional state vector
\begin{equation}
 \vl \psi (t) \rangle = \sum_{n=0}^N c_n \vl n \rangle ,
\end{equation}
where $|n\rangle = |0\rangle_A|00\ldots 1_n \ldots 00\rangle$ denotes
the state in which the $n$th qubit of the chain is the only one in
state 1.  In order to achieve a compact notation, we have introduced
the state $|0\rangle = |1\rangle_A|0\ldots0\rangle$ for which all
qubits are in the ground state, while Alice' state is excited.

The main requirement for our protocol is that Alice' qubit remains
entangled with the qubit that propagates to Bob or to Charlie,
respectively.  We measure this property by the concurrence
$C=\max\{\lambda_1-\lambda_2-\lambda_3-\lambda_4,0\}$
\cite{Wootters1998}.  The $\lambda$'s are the ordered square roots of
the eigenvalues of the matrix $\rho(\sigma_y^1 \otimes \sigma_y^2)
\rho^* (\sigma_y^1 \otimes\sigma_y^2)$ with $\rho$ being the reduced
density matrix of the considered qubit pair.
In our case, the relevant reduced density matrix is that of Alice'
qubit and qubit $n$ of the chain.  Tracing out all other qubits, we
obtain in the basis $\{ \vl 0_A 0_n \rangle,\vl 0_A 1_n \rangle,\vl
1_A0_n \rangle,\vl 1_A1_n \rangle\}$ the expression
\begin{equation}
\rho_{A,n} =
\begin{pmatrix}
\sum_{i\neq 0,n} |c_i|^2 & 0 &0, &0
\\
0 & |c_0|^2 & c_n^* c_0 &  0
\\
0 & c_0^* c_n & |c_n|^2 & 0
\\
0 & 0 & 0 &0
\end{pmatrix}
\end{equation}
for which the concurrence reads
\begin{equation}
\label{C_An}
C_{A,n} = 2 \vl c_0 c_n \vl.
\end{equation}

\begin{figure}[tb]
\centerline{\includegraphics[width=.95\columnwidth]{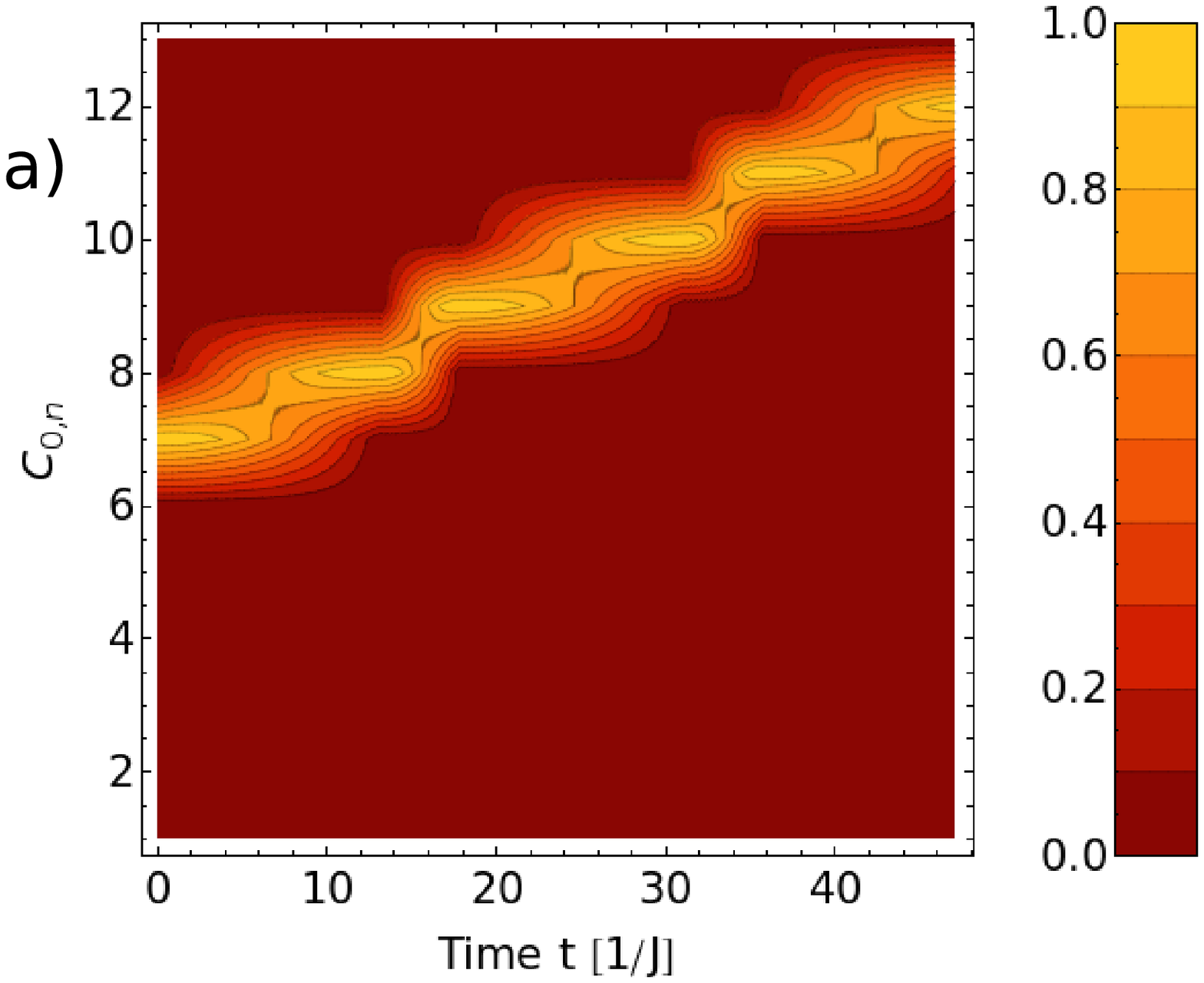}}
\centerline{\includegraphics[width=8.5cm]{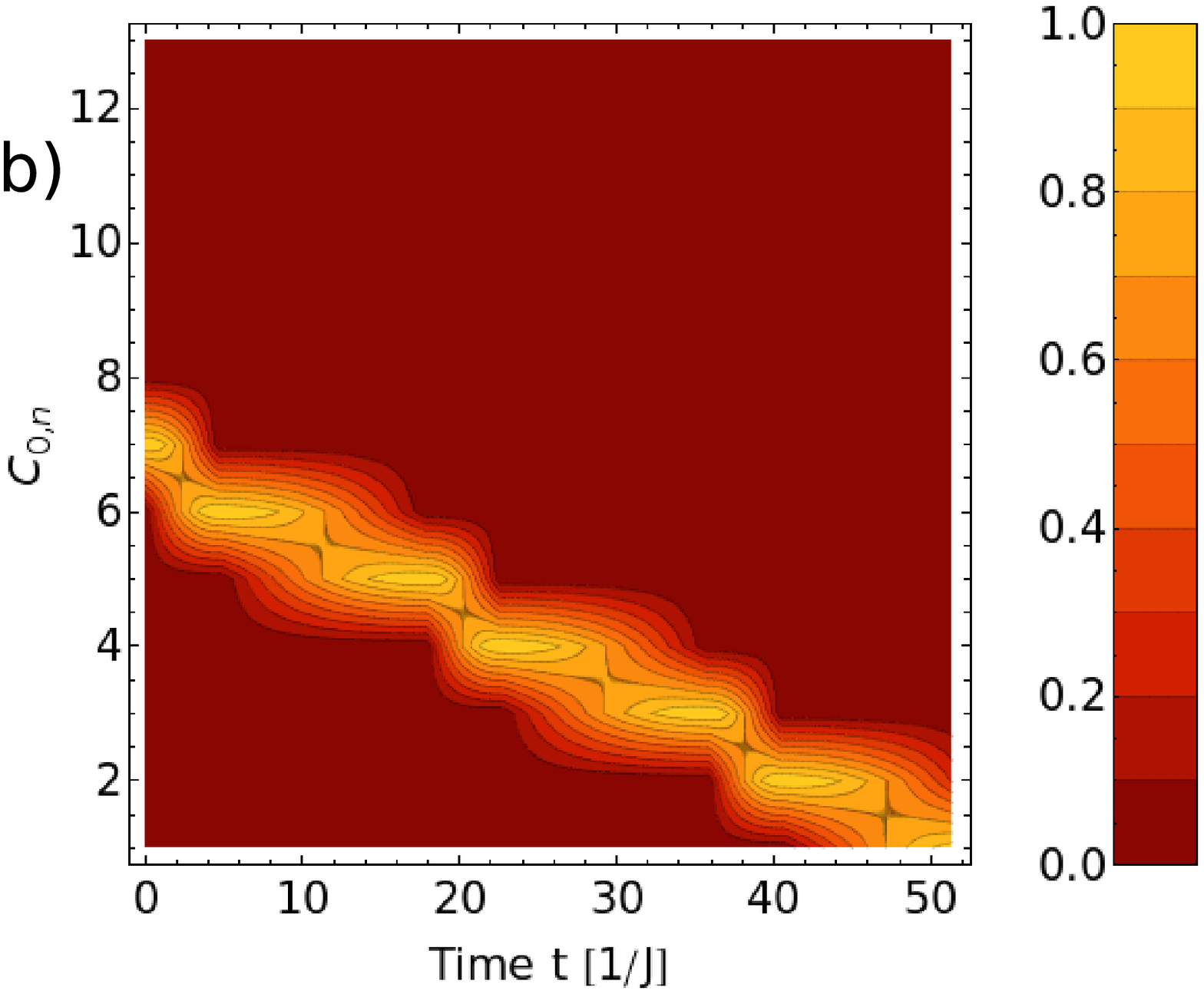}}
\caption{(a) Time evolution of the concurrence between Alice' qubit and
qubit $n$ of the vertical chain with length $N=13$ for the parameters
$\omega=10J$, $b=J$, $\Lambda_1=3J$, $\Lambda_2=1.5J$.  Alice' qubit
is initially entangled with the qubit at $n_\text{node}=7$.  The onset
of the driving field \eqref{coseno} is such that initially transfer to
Charlie is suppressed.
(b) Same as panel (a) but for an arrival time $t = T_1$ at the node,
such that the entanglement propagates to Charlie.
}
\label{fig2}
\end{figure}

If Alice transfers her qubit directly to the node, i.e., to the
central qubit $n_\text{node}$ of the vertical chain, we can start our
numerical calculation with the initial state
\begin{equation}
\ket{\psi(0)} = \frac{1}{\sqrt{2}}(\ket{0}+\ket{n_\text{node}}) .
\end{equation}
Figure~\ref{fig2} shows the corresponding time evolution of the
concurrence $C_{A,n}$.  It demonstrates that for a proper onset of the
driving field \eqref{coseno}, the entanglement propagates solely to
Bob who finally receives a state that is perfectly entangled with
Alice' state.  For a driving field starting with the larger of the
two amplitudes, the entanglement propagates equally
perfectly to Charlie, see Fig.~\ref{fig:T}.  Thus, one can route the
entanglement to a particular end of the vertical chain by choosing a
proper initial time of the driving field.

\begin{figure}[tb]
 \includegraphics[width=.95\columnwidth]{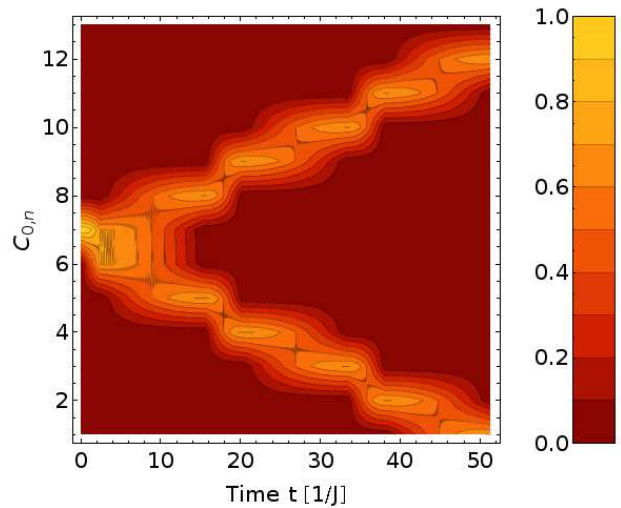}
 \caption{Same as Fig.~\ref{fig2} for the arrival time $t=T_1/2$ at
the node, such that the travelling qubit is split.}
\label{fig4}
\end{figure}

It is also possible that both Bob and Charlie receive a state that is
entangled with Alice' state, as can be appreciated in Fig.~\ref{fig4}.
There the travelling qubit arrives with a delay of $t=T_1/2$, such
that the routing towards Bob is only half completed at the time the
upward channel is closed and routing to Charlie begins.

\begin{figure}[tb]
\centerline{\includegraphics[width=.95\columnwidth]{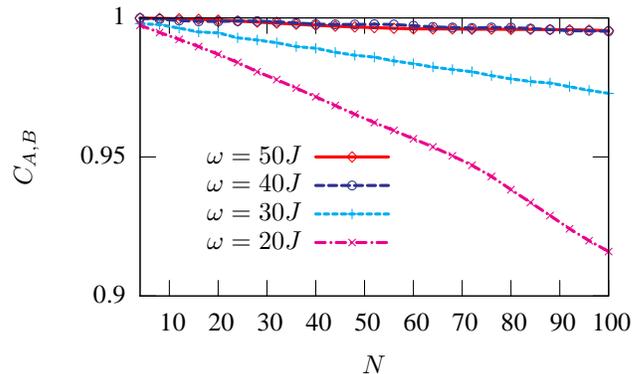}}
\caption{Length dependence of the final concurrence for
chains of various frequencies.  The differences of the energy splittings
are $\Lambda_1 =100 J$ and $\Lambda_2=59.02 J$, corresponding to the
ratio $\Lambda_2/\Lambda_1 = 0.5902$ which provides the optimum transfer velocity.
}
\label{concW}
\end{figure}

Our intuition for the dynamics of the quantum router  is based on
the derivation of an effective time-independent Hamiltonian derived
within high-frequency approximation.  
In the following we
numerically test the results of our high-frequency approximation for
finite driving frequencies.
In Fig.~\ref{concW} we plot the final concurrence $C_{A,B}$ of
Alice' qubit and the qubit that Bob receives, for different driving
frequencies.  We find that the protocol works, provided that the
driving frequency is sufficiently large.  For chains of length up to
$N=100$, the final entanglement is quite large, e.g., $C_{A,B}\geq
0.97$ for $\omega\geq 30J$.  For driving frequencies $\omega\lesssim
20J$, the concurrence already assumes a significantly lower value.

In Fig.~\ref{concW}, we used a driving field with the amplitude ratio
$\Lambda_1/\Lambda_2 = 0.5902$ for which the entanglement transfer is
as fast as possible for a given value of $J$.  Then the operation time
and, thus, the total coherence loss is kept to a minimum, see
Sec.~\ref{sec:decoherence}.  In order to obtain the optimized
amplitude ratio, one has to consider that during a full cycle of the
driving field, the entanglement is transported over two sites.  This
takes the time
\begin{equation}
\label{T1T2}
T_1 + T_2
= 
\frac{\pi}{2|J_0(\Lambda_1/\Lambda_2)|}
       +\frac{\pi}{2|J_0(\Lambda_2/\Lambda_1)|} ,
\end{equation}
which possesses a minimum for the mentioned ratio and for the inverse
ratio as well.

\subsection{Quantum state transfer}

The spin chain can also be used for quantum state transfer, such that
a classical communication channel becomes dispensable \cite{bose03}.
For investigating state transfer, we assume that Alice owns the state
\begin{equation}
\label{st-A}
\vl \psi \rangle_{A} =
 \alpha \vl 0 \rangle + \beta \vl 1 \rangle \; ,
\end{equation}
which she likes to transfer to Bob.  In doing so, she initializes the
chain with the first qubit in the target state, while the other qubits
of the horizontal chain are set to the ground state.  For an
excitation-preserving spin chain such as the one defined by the
Hamiltonian \eqref{iso-gral}, it can be shown that the unitary
evolution of the chain provides Bob with the state \cite{Bose2006}
\begin{equation}
\label{rhobob}
\rho_B =
\begin{pmatrix}
1 - \vl \beta \vl^2  \vl c_B \vl & \alpha \beta^* c_B
\\
\alpha^* \beta c_B & \vl \beta \vl^2  \vl c_B \vl
\end{pmatrix}
\end{equation}
written in the basis $\{\ket {0}, \ket{1}\}$.  Perfect state transfer
corresponds to $|c_B|=1$.  For our driven spin chain with different
local qubit splittings and ac fields, each transfer from one qubit to
the next contributes a particular phase shift.  Thus, Bob will receive
a state with $c_B=\vl c_B \vl \e^{\iu \theta}$.  In order to restore
the full state \eqref{st-A}, one has to know the total phase
shift $\theta$, which requires full knowledge of the transfer process,
i.e., one has to calibrate the chain before use.  Subsequently, a
fixed unique phase shift has to be applied to every signal coming
from the hub.  Such hub together with a data sheet that contains the
phase shifts could represent a useful piece of hardware.

Finally, the quality of the state transfer can be quantified with
the fidelity
\begin{equation}
\label{fidelity}
 F = \vl \langle \psi_{A} \vl \rho_{B} \vl \psi_{A} \rangle  \vl \: ,
\end{equation}
which in our case reads
\begin{equation}
F=1-(1-C_{A,B})|\beta|^2\left[1-C_{A,B}(1-2|\beta|^2)\right] \; ,
\label{FtoC}
\end{equation}
after the phase has been rotated back.  This means that the fidelity
directly relates to the concurrence, $C_{A,B}$ in Eq. ~\eqref{C_An}, of the entanglement
router.

\section{Physical implementations}
\label{sec:implementation}

A highly accurate and stable confinement of ions in traps has been
achieved in the last decade, together with the possibility of working in
the ground state of the ion motion \cite{Roos2000a, wineland}. This
enables the realization of quantum chains that consist of harmonic
oscillators with a nearest-neighbor coupling. The physical origin of
the coupling can be Coulomb interaction between the confined particles
or a capacitive coupling of the ions residing in neighboring wells;
see e.g.\ Refs.~\cite{Ciaramicoli2003, Stahl2005}.
For a comprehensive review of according control and readout
techniques, see Ref.~\cite{wineland}.

The Hamiltonian for the excitations in Eq.~(\ref{excit}), with a
general onsite Zeeman field $h_n(t)$ can also be implemented by a
chain of harmonic oscillators. The particular form of the interaction
is typically obtained by coupling the spatial coordinates of
neighboring oscillators [see Eq.~(\ref{osci})], as is the case for
Coulomb interaction between charged particles in dipole approximation.
Note however, that the resulting Hamiltonian contains terms that
create and annihilate phonon pairs.  These terms vanish within
rotating-wave approximation, which means that a mapping to our
spin-chain model \eqref{iso-gral} is possible only if this
approximation holds, i.e., if the coupling is much weaker than the
time average of an applied ``Zeeman'' field, $J \ll h_0$, where here
$h_n(t) = h_0 +b_n(t)$.  Notice that the presence of the constant
``Zeeman'' field $h_0$ merely rescales the energies and will not
affect the router otherwise.  Experimentally, this field can be
realized by an ac voltage with non-zero mean, such that
$h_n(t)=2\hbar\omega_n(t)$ with the frequency $\omega_n(t)$
proportional to the square root of the applied voltage.

As a particular example, we consider singly trapped ions in planar
Penning traps, in which neighboring ions couple capacitively through a conducting wire.
Neglecting resonant absorption in the wire, the coupling for a
typical trap size of 1mm will be roughly 100 times lower than the
onsite potential \cite{Stahl2005}. For even smaller traps of size
0.1mm, these quantities still differ by a factor of 10.  Thus,
engineering the coupling strength essentially means choosing the
correct trap size. For planar traps, the ion can be moved further away
from the electrodes, which reduces the coupling even more.  In a
linear array of trapped ions such as the one of
Ref.~\cite{Ciaramicoli2003}, the atoms experience Coulomb interaction.
In both cases, the ratio between interaction strength and onsite
potential $\hbar\omega$ is proportional to $q^2/(\pi \epsilon_0
m\omega^2 d^3)$. The difference in strength between both approaches is
a form factor, given by the fact that the wire-mediated coupling has a
given efficiency when transmitting Coulomb interaction.  In the case
of direct Coulomb interaction, the coupling is slightly stronger, but
can be reduced by pushing the ions far apart from each other, using
that fact that $J$ is reciprocal to the typical distance between the
ions.

It must be noted here that for Penning traps, all parameters basically
depend on two scales, namely the trap size and the voltage.  The
motional frequency is proportional to $\sqrt{V}/d$, with $V$ the
applied voltage and $d$ the size of the trap. Manipulating both scales
allows one to achieve the desired parameter values.

\subsection{Robustness under realistic conditions}

In an experiment, one usually faces additional difficulties not
captured by a Hamiltonian with perfectly stable parameters.  Two such
difficulties come to mind, namely imperfections in the fabrication
process and the unavoidable influence of an environment that causes
dissipation and decoherence.

\subsubsection{Fabrication uncertainties}

An essential ingredient to our entanglement distribution protocol is
the action of the onsite ac fields $h_n =
b_n(\xi_0\omega/\Lambda_i)\sin\omega t$ during times $T_i =
\pi/J_{\text{eff},i}$ with $J_{\text{eff},1(2)} = J
J_0(\xi_0\Lambda_{1(2)}/\Lambda_{2(1)})$, [see Eq.~(\ref{coseno}) and
Fig.~\ref{driving}]. This implies two critical experimental
requirements.  First, the driving amplitudes have to match with good
precision the first zero of the Bessel function and, second, the field
amplitudes have to be switched after a time $T_{1,2}$.

Trap diameters
in the sub-millimeter range, correspond to motional frequencies of the
order GHz, with a $J\sim100$MHz, such that the rotating wave
approximation applies.  Moreover, a driving frequency of the order
0.1\,GHZ is available with current technology, while commercially
available function generators at those frequencies can have an
accuracy of several $10^9$ samples per second. Thus the switching
times can be adjusted with precision $\epsilon_T = 10^{-3}/J$.  The
uncertainty in $J$ mostly comes from measuring the distance
between trapped ions (typically by fluorescence).  Assuming an
accuracy of $10\%$, we obtain the error $\epsilon_J=10^{-7}J$, which
corresponds to $\epsilon_T=10^{-7}/J$. Thus, the relevant
restriction is the mentioned uncertainty in the
function generator, such that $\epsilon_T = 10^{-3}/J$.  Concerning
the amplitudes, we assume a relative error in the applied
confining voltages of the order of $10^{-6}$, which yields an
amplitude uncertainty of the order of $10^{-7}J$.

For a numerical simulation of these errors, we choose for each coupling
matrix element $J_i$ a random value from the interval $[J-\epsilon_J,
J+\epsilon_J]$ with equal distribution.  The interval width
$2\epsilon_J$ is determined by the experimental uncertainties
discussed above.  Accordingly, we select a field amplitude from
the interval $[b_n - \epsilon_b, b_n + \epsilon_b]$.  In the
following, we denote ensemble averages of such realizations by an
overbar.

We already emphasized in Sec.~\ref{sec:cdt} that the protocol for
entanglement transfer is closely related to the one for state
transfer.  The main difference is that the latter requires knowledge
of the extra phase $\theta$.  It has to be gauged or determined from
theoretical considerations, such that Bob is able to compensate this
phase by a local transformation.  When the system parameters
acquire a random component, the extra phase becomes random as well.

In principle, the additional random phase can be compensated by local
transformations.  Thus, it cannot influence the entanglement.
Numerically, we have tested that the experimental uncertainties mainly
affect this phase, making the concurrence  fairly independent of this
type of errors.  Therefore we focus on the impact of the fabrication
uncertainties on the state transfer.  Assuming that we can
gauge our apparatus, we know that the actual value of $\theta =
\bar\theta+\Delta_\theta$ deviates from the average by
$\Delta_\theta$.  Notice that $\bar \theta$ can be obtained just by
averaging over many realizations, as we discuss below. Then the fidelity
\eqref{fidelity} becomes [cf.\ Eq.~\eqref{FtoC}]
\begin{equation}
\begin{split}
F ={} & 1 -  \vl \beta \vl^2 (1 -2 C_{A,B} \cos\Delta_\theta+ C_{A,B}^2)
\\
&+ 2 \vl \beta \vl^4   C_{A,B}(C_{A,B}-\cos\Delta_\theta) \; .
\end{split}
\label{F-fabe}
\end{equation}
Thus, all the possible errors are captured by the phase deviation
$\theta = \bar \theta + \Delta_\theta$.

\begin{figure}[tb]
\begin{center}
\includegraphics[width=.95\columnwidth]{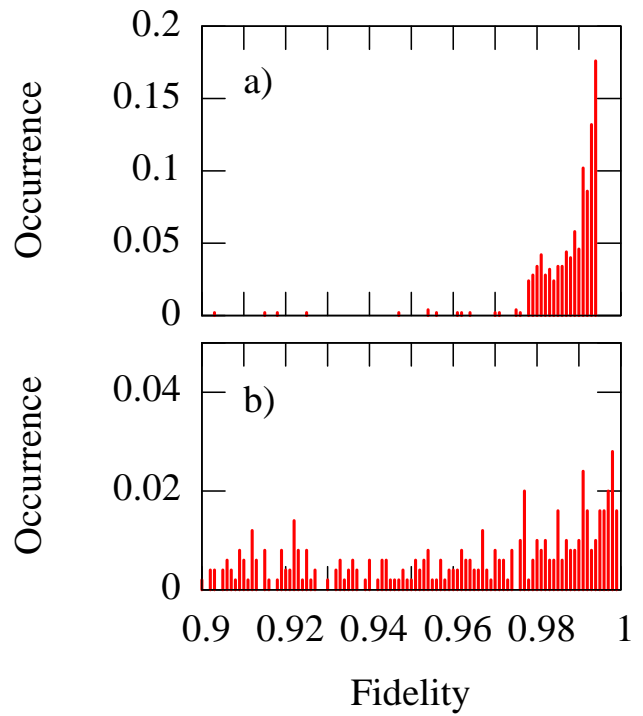}
\end{center}
\caption{Fidelity distribution for the initial state $\ket {\psi}_{A}=
  (\ket{0}+\ket{1})/\sqrt{2} $ in a chain with length $N=40$ and the
  driving frequencies (a) $\omega=30$ and (b) $\omega=40$.
  The qubits splittings are $\Lambda_1 =100 J$ and
  $\Lambda_2=59.02 J$, while the parameter uncertainties read
  $\epsilon_J= \epsilon_b=10^{-7}$ and $\epsilon_T=10^ {-3}$.  The
  data have been obtained from $500$ realizations.  The resulting
  standard deviations of the concurrence are $0.01$ and $0.1$,
  respectively.}
\label{fig:fidelity}
\end{figure}

Figure~\ref{fig:fidelity} shows the frequentness of the accordingly
reduced values for the fidelity.  We have computed the
fidelity~\eqref{fidelity} for 500 realizations of the spin chain, and
plotted a histogram with bin size $10^{-3}$.  Obviously, the driving
frequency $\omega=30J$ yields in most cases a fidelity $F>0.98$
[panel (a)].  For a larger driving frequency [panel (b)], the fidelity
distribution is still peaked at a value larger than $0.99$.  The
standard deviation, however, is quite large, such that significantly
smaller fidelities become rather likely.  Thus, fabrication
uncertainties impose that the driving frequency should not exceed
$30J$.  On the other hand, we have seen in Sec.~\ref{sec:routing} that
this value in fact also represents a lower bound for the validity of the
rotating-wave approximation within which the effective coupling
strength \eqref{Jeff} has been derived.  Thus, we conclude that
the results are best for a driving frequency of the order $\omega =
30J$.

\subsubsection{Dissipation and decoherence}
\label{sec:decoherence}

Still we have to consider the main obstacle for any quantum
information task, which is dissipation and decoherence caused by weak
but unavoidable coupling of the processor to a typically huge number
of uncontrollable degrees of freedom.  Irrespective of its physical
nature, the environment is usually modeled as a bath of harmonic
oscillators, where each oscillator couples via its position operator
to a system coordinate \cite{magalinskii1959, Caldeira1983, GrifoniPR,
Chaos2005}.  Here we assume that each qubit $n$ undergoes pure and
independent dephasing, i.e., that it its coordinate $\sigma_n^z$
couples to a separate bath.  

For weak dissipation, one can eliminate the bath within second-order
perturbation theory and derive a Bloch-Redfield master equation for
the reduced density operator $\rho$ of the qubits \cite{Blum1996a}. 
Considering, only phase noise on the traps the equation takes a Lindblad form 
\cite{Lindblad1976a}, 
\begin{equation}
\label{qme}
\dot{\rho}
= -\frac{\iu}{\hbar}[H,\rho]
  +\frac{\gamma}{2} \sum_{n=1}^N (\sigma_n^z\rho\sigma_n^z-\rho),
\end{equation}
where we have assumed that the effective decoherence rate $\gamma$ is
the same for all qubits.  Remarkably, such a master equation
describes, apart from a transient, the decoherence process quite
accurately \cite{Doll2008}.  For qubit arrays implemented with calcium
ions confined in a Paul trap, the dephasing rate is $\gamma \sim
1\,\mathrm{kHz}$, while the qubit splitting is several MHz.  The
heating rate, which would include relaxation terms in the master
equation, is $190$ times smaller \cite{Roos1999}.   
This justifies the phase noise model used here.  We will use this
numbers for a rather conservative estimate and assume
$\gamma\sim10^{-4}J$, where $J\sim10\,\mathrm{MHz}$.

For qubit chains under the influence of phase noise, the entanglement
decay is determined by the coherence loss, and one finds that the
concurrence decays exponentially with a rate $2\gamma$, i.e.,
$C_{A,B}(t) \propto \exp(-2\gamma t)$ \cite{Doll2007a}.  This can be
verified easily by considering the action of the dissipative kernel in
the master equation on the off-diagonal density matrix elements
$\rho_{nn'}$.  The
entanglement propagation from the middle of the chain to Bob's end
takes the time $\tau_{AB} = (T_1+T_2)N/4$ [Recall that our ratchet
mechanism transports signals during a time $T_1+T_2$ over two sites;
see derivation of Eq.~\eqref{T1T2}].  Thus, we expect that decoherence
reduces the finally achieved entanglement according to
\begin{equation}
\label{dec-estimate}
 C_{A,B} = C_{A,B} (\gamma{=}0) \e^{- 2 \gamma \tau_{AB}} .
\end{equation}

\begin{figure}[tb]
\begin{center}
\includegraphics[width=.95\columnwidth]{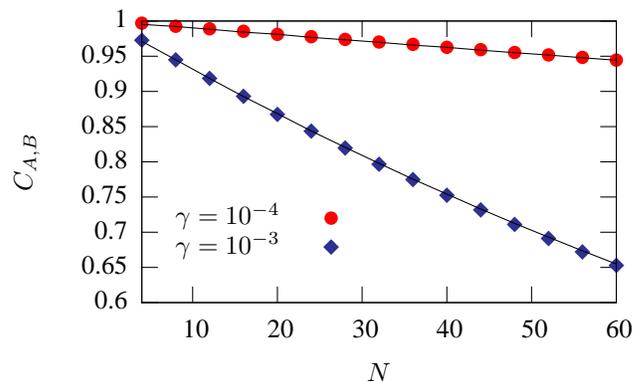}
\end{center}
\caption{Concurrence decay as a function of the chain length for two
different decoherence rates $\gamma$ (symbols) and $\omega=30$.  The
solid lines mark the estimate~\eqref{dec-estimate}.  All other
parameters are as in Fig.~\ref{fig:fidelity}.
}
\label{fig:concG}
\end{figure}%
We numerically computed the time evolution described by the master
equation~\eqref{qme}.  The results shown in Fig.~\ref{fig:concG}
confirm our expectations for the exponential decay.  Moreover, they
demonstrate that a decoherence rate of $\gamma = 10^{-4}J$ reduces the
concurrence in a chain of length $N=60$ from
$C_{A,B}(\gamma{=}0) =0.98$ in the absence of decoherence to a value
$C_{A,B} = 0.95$.  This is still sufficiently large for most possible
applications.  For a larger decoherence rate, $\gamma = 10^{-3}J$,
however, the concurrence drops significantly.  For example in a chain of
length $N=60$, we have the already quite small value $C_{A,B} = 0.65$.
Thus, we can conclude that for chains of this length,
transmission of sufficient entanglement is only possible for
decoherence rates that fulfill $\gamma \lesssim 10^{-4}J$. 
For stronger decoherence, a hub with shorter chains is nevertheless
possible, in the extreme limit even one with branches that consist of
just one qubit.

We close our analysis of fabrication uncertainties and decoherence by
addressing also the fidelity of the state transfer protocol.  In the
presence of decoherence, it is no longer possible to establish an
exact relation between the fidelity and the concurrence [cf.\
Eq.~\eqref{FtoC}].  
The reason for this is that state transfer is limited by the dephasing
of \textit{individual} qubits, while the entanglement decay is
influenced by the coherence between \textit{different} qubits.
The state transfer fidelity to Bob can still be expressed in terms
of Bob's reduced density matrix~\eqref{rhobob} and now reads
\begin{equation}
\label{fidelityG}
 F = 1 - \rho_{11} +\vl \beta \vl ^2 (2 \rho_{11} -1) + 2 \vl
\rho_{10} \vl \mathop{\mathrm{Re}} ( \alpha \beta \e^{\iu \Delta_\theta}) ,
\end{equation}
which implies that again the average phase shift $\bar \theta$ plays a
role; cf.\ Eq.~\eqref{F-fabe}.

We find numerically, that the final population of Bob's qubit, which
is described by the density matrix element $\rho_{BB}$, is practically
independent of both the decoherence strength $\gamma$ and the
fabrication uncertainties.  Without these influences, the state
transfer protocol works almost perfectly, such that we can approximate
this density matrix element by $\rho_{BB} = |\beta|^2$.  Moreover, we
already argued that off-diagonal matrix elements decay as $\rho_{B0} =
\alpha \beta^* \exp ( - \gamma \tau_{AB})$ and, thus, the finally
achieved fidelity can be well aproximated by
\begin{equation}
F = 1 - 2 (\vl \beta \vl ^2-\vl \beta \vl ^4 )[ 1 - \cos (
\Delta_\theta) \e^{- \gamma \tau_{AB}}] \; .
\label{app}
\end{equation}
Thus, within some reasonable approximations, we again find for the
fidelity a closed expression that is a function of the initial state.
In fact, for $\gamma=0$ and $C_{AB}(\gamma=0) = 1$,
expression~\eqref{app} becomes identical with the fidelity~\eqref{F-fabe}
for an ideal chain.  This also demonstrates that the fidelity is
significantly affected by both fabrication uncertainties and
decoherence, expressed in Eq.~\eqref{app} by the random phase shift
$\Delta_\theta$ and the exponentially decaying factor, respectively.

\begin{figure}[tb]
\begin{center}
\includegraphics[width=.95\columnwidth]{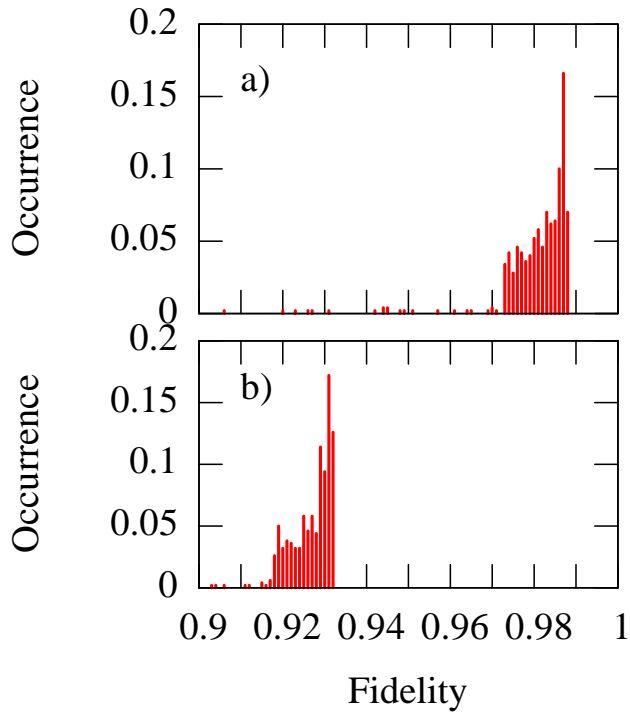}
\end{center}
\caption{Fidelity spread for a chain of length $N=40$ with the initial
state $\ket{\psi}_A= ( \ket{0} + \ket{1} )/\sqrt{2}$.  The driving
frequency is $\omega=30$ and dissipation strengths
 $\gamma=10^{-3}J$ (a)  and  $\gamma=10^{-4}J$ (b). The histogram is
obtained from $500$ realizations yielding
a standard deviation of  $0.01$ in both of them.  All other parameters are as in
Figs.~\ref{fig:fidelity}, \ref{fig:concG}. 
}
\label{fig:fidelity-deco}
\end{figure}%
It remains to numerically corroborate the quality of the approximate
result~\eqref{app}.  Figure~\ref{fig:fidelity-deco} depicts the
numerical results for a chain of length $N=40$.  
It reveals that the maximum fidelity in the histogram coincides with making $\Delta_\theta = 0$ in 
\eqref{app}, namely $F=0.99$ for $\gamma=10^{-4}J$  and $F=0.93$ for $\gamma=10^{-3}J$.
Further the data spread, due to the fabrication errors, is pretty the same that in the case without decoherece [cf. Fig. \ref{fig:fidelity}].   This  underlines that state
transfer with fidelity $F \gtrsim 0.97$ is achievable with the
experimental realization discussed above, provided the dissipation
rate does not exceed the value $\gamma = 10^{-4}J$.

\section{Discussion and conclusion}

We have proposed a quantum router that can be employed for steering
one qubit of an entangled pair from Alice to either Bob or Charlie.  A
second, related application is quantum state transfer.  With the
experimental realization with ion traps in mind, we have presented
analytical and numerical results that underline the feasibility of our scheme.

The underlying physics is coherent destruction of tunneling between
two neighboring qubits.  This can be achieved by ac fields with a
certain ratio between driving amplitude and frequency.  Since the
effective amplitude depends on the difference of the qubits' Zeeman
energies, a proper sequence of these energies temporarily suppresses
for each qubit the interaction with exactly one of its two neighbors.
After half a tunnel cycle, the field amplitude is switched, such that
the suppression is active for the connection to the other neighbor.
As a consequence, the qubit is transported into one particular
direction.  A main benefit of this mechanism is that it does not rely
on the local control of interaction parameters, but only on the less
demanding application of proper ac fields.

In the T-shaped configuration under investigation, one partner of an
entangled pair is first routed to the node that connects Alice with
Bob and Charlie.  When the qubit arrives at the node, the further
propagation depends on which branch is temporarily closed at that very
moment.  In particular, we demonstrated that for a proper phase of the
driving field, the qubit is reliably routed towards the one or the
other direction.  The achievable entanglement increases with the
driving frequency.  For an idealized chain of 100 qubits with
realistic parameters, the corresponding concurrence may assume values
of the order 0.98.

In an experiment, however, the setup and the operation will not be
ideal.  Therefore, we have investigated two limiting influences,
namely fabrication imperfections of the desired splitting sequence and
the influence of external degrees of freedom which lead to
decoherence.
The former reduce the achievable fidelity, in particular for high
driving frequencies, i.e., in the regime in which the idealized model
works almost perfectly.  The numerical simulation revealed the
existence of an intermediate frequency regime,  in which realistic fabrication
errors  do not significantly reduce the achievable
fidelity.

The second limitation leads to the ubiquitous decoherence which
generally represents the main obstacle for the implementation of
quantum information schemes. 
It turned out that dephasing does not limit the entanglement or state transfer
as long as its rate is roughly three orders of magnitude smaller than
the typical qubit splitting.

In summary, we have demonstrated that the proposed quantum router
should work under realistic conditions such as fabrication
uncertainties and decoherence.  The implementation is feasible
not only with ion chains, but with practically all interacting qubit
arrays in a T-shaped configuration.  Thus our quantum router may
represent an essential building block in future quantum information
networks.

\begin{acknowledgments}
We gratefully acknowledge financial support by the German Excellence
Initiative via the ``Nanosystems Initiative Munich (NIM)''.
This work has been supported by DFG through SFB 484 and SFB 631.
\end{acknowledgments}

\bibliography{galve_zueco,decoherence}

\begin{thebibliography}{45}
\expandafter\ifx\csname natexlab\endcsname\relax\def\natexlab#1{#1}\fi
\expandafter\ifx\csname bibnamefont\endcsname\relax
  \def\bibnamefont#1{#1}\fi
\expandafter\ifx\csname bibfnamefont\endcsname\relax
  \def\bibfnamefont#1{#1}\fi
\expandafter\ifx\csname citenamefont\endcsname\relax
  \def\citenamefont#1{#1}\fi
\expandafter\ifx\csname url\endcsname\relax
  \def\url#1{\texttt{#1}}\fi
\expandafter\ifx\csname urlprefix\endcsname\relax\def\urlprefix{URL }\fi
\providecommand{\bibinfo}[2]{#2}
\providecommand{\eprint}[2][]{\url{#2}}

\bibitem[{\citenamefont{Deutsch}(1985)}]{Deutsch1985}
\bibinfo{author}{\bibfnamefont{D.}~\bibnamefont{Deutsch}},
  \bibinfo{journal}{Proc. R. Soc. A} \textbf{\bibinfo{volume}{400}},
  \bibinfo{pages}{97} (\bibinfo{year}{1985}).

\bibitem[{\citenamefont{{Ursin} et~al.}(2007)\citenamefont{{Ursin},
  {Tiefenbacher}, {Schmitt-Manderbach}, {Weier}, {Scheidl}, {Lindenthal},
  {Blauensteiner}, {Jennewein}, {Perdigues}, {Trojek} et~al.}}]{Zeillinger2007}
\bibinfo{author}{\bibfnamefont{R.}~\bibnamefont{{Ursin}}},
  \bibinfo{author}{\bibfnamefont{F.}~\bibnamefont{{Tiefenbacher}}},
  \bibinfo{author}{\bibfnamefont{T.}~\bibnamefont{{Schmitt-Manderbach}}},
  \bibinfo{author}{\bibfnamefont{H.}~\bibnamefont{{Weier}}},
  \bibinfo{author}{\bibfnamefont{T.}~\bibnamefont{{Scheidl}}},
  \bibinfo{author}{\bibfnamefont{M.}~\bibnamefont{{Lindenthal}}},
  \bibinfo{author}{\bibfnamefont{B.}~\bibnamefont{{Blauensteiner}}},
  \bibinfo{author}{\bibfnamefont{T.}~\bibnamefont{{Jennewein}}},
  \bibinfo{author}{\bibfnamefont{J.}~\bibnamefont{{Perdigues}}},
  \bibinfo{author}{\bibfnamefont{P.}~\bibnamefont{{Trojek}}},
  \bibnamefont{et~al.}, \bibinfo{journal}{Nature Phys.}
  \textbf{\bibinfo{volume}{3}}, \bibinfo{pages}{481} (\bibinfo{year}{2007}).

\bibitem[{\citenamefont{{Bose}}(2003)}]{bose03}
\bibinfo{author}{\bibfnamefont{S.}~\bibnamefont{{Bose}}},
  \bibinfo{journal}{Phys. Rev. Lett.} \textbf{\bibinfo{volume}{91}},
  \bibinfo{pages}{207901} (\bibinfo{year}{2003}).

\bibitem[{\citenamefont{{Bose}}(2007)}]{Bose2006}
\bibinfo{author}{\bibfnamefont{S.}~\bibnamefont{{Bose}}},
  \bibinfo{journal}{Contemp. Phys.} \textbf{\bibinfo{volume}{48}},
  \bibinfo{pages}{13} (\bibinfo{year}{2007}).

\bibitem[{\citenamefont{{Christandl} et~al.}(2004)\citenamefont{{Christandl},
  {Datta}, {Ekert}, and {Landahl}}}]{Christandl2004}
\bibinfo{author}{\bibfnamefont{M.}~\bibnamefont{{Christandl}}},
  \bibinfo{author}{\bibfnamefont{N.}~\bibnamefont{{Datta}}},
  \bibinfo{author}{\bibfnamefont{A.}~\bibnamefont{{Ekert}}}, \bibnamefont{and}
  \bibinfo{author}{\bibfnamefont{A.~J.} \bibnamefont{{Landahl}}},
  \bibinfo{journal}{Phys. Rev. Lett.} \textbf{\bibinfo{volume}{92}},
  \bibinfo{pages}{187902} (\bibinfo{year}{2004}).

\bibitem[{\citenamefont{{Nikolopoulos}
  et~al.}(2004)\citenamefont{{Nikolopoulos}, {Petrosyan}, and
  {Lambropoulos}}}]{Nikolopoulos2004}
\bibinfo{author}{\bibfnamefont{G.~M.} \bibnamefont{{Nikolopoulos}}},
  \bibinfo{author}{\bibfnamefont{D.}~\bibnamefont{{Petrosyan}}},
  \bibnamefont{and}
  \bibinfo{author}{\bibfnamefont{P.}~\bibnamefont{{Lambropoulos}}},
  \bibinfo{journal}{Europhys. Lett.} \textbf{\bibinfo{volume}{65}},
  \bibinfo{pages}{297} (\bibinfo{year}{2004}).

\bibitem[{\citenamefont{{di Franco}
  et~al.}(2008{\natexlab{a}})\citenamefont{{di Franco}, {Paternostro}, and
  {Kim}}}]{diFranco2008}
\bibinfo{author}{\bibfnamefont{C.}~\bibnamefont{{di Franco}}},
  \bibinfo{author}{\bibfnamefont{M.}~\bibnamefont{{Paternostro}}},
  \bibnamefont{and} \bibinfo{author}{\bibfnamefont{M.~S.} \bibnamefont{{Kim}}},
  \bibinfo{journal}{Phys. Rev. Lett.} \textbf{\bibinfo{volume}{101}},
  \bibinfo{pages}{230502} (\bibinfo{year}{2008}{\natexlab{a}}).

\bibitem[{\citenamefont{Wu et~al.}(2009)\citenamefont{Wu, Miranowicz, Wang,
  Liu, and Nori}}]{wu2009}
\bibinfo{author}{\bibfnamefont{L.-A.} \bibnamefont{Wu}},
  \bibinfo{author}{\bibfnamefont{A.}~\bibnamefont{Miranowicz}},
  \bibinfo{author}{\bibfnamefont{X.}~\bibnamefont{Wang}},
  \bibinfo{author}{\bibfnamefont{Y.-X.} \bibnamefont{Liu}}, \bibnamefont{and}
  \bibinfo{author}{\bibfnamefont{F.}~\bibnamefont{Nori}}
  (\bibinfo{year}{2009}), \eprint{arXiv:0902.3564}.

\bibitem[{\citenamefont{{di Franco}
  et~al.}(2008{\natexlab{b}})\citenamefont{{di Franco}, {Paternostro},
  {Tsomokos}, and {Huelga}}}]{diFranco2008b}
\bibinfo{author}{\bibfnamefont{C.}~\bibnamefont{{di Franco}}},
  \bibinfo{author}{\bibfnamefont{M.}~\bibnamefont{{Paternostro}}},
  \bibinfo{author}{\bibfnamefont{D.~I.} \bibnamefont{{Tsomokos}}},
  \bibnamefont{and} \bibinfo{author}{\bibfnamefont{S.~F.}
  \bibnamefont{{Huelga}}}, \bibinfo{journal}{\pra}
  \textbf{\bibinfo{volume}{77}}, \bibinfo{pages}{062337}
  (\bibinfo{year}{2008}{\natexlab{b}}).

\bibitem[{\citenamefont{{Romero-Isart} and
  {Garc{\'{\i}}a-Ripoll}}(2007)}]{romeroisart2007}
\bibinfo{author}{\bibfnamefont{O.}~\bibnamefont{{Romero-Isart}}}
  \bibnamefont{and} \bibinfo{author}{\bibfnamefont{J.~J.}
  \bibnamefont{{Garc{\'{\i}}a-Ripoll}}}, \bibinfo{journal}{\pra}
  \textbf{\bibinfo{volume}{76}}, \bibinfo{pages}{052304}
  (\bibinfo{year}{2007}).

\bibitem[{\citenamefont{You and Nori}(2005)}]{You2005}
\bibinfo{author}{\bibfnamefont{J.}~\bibnamefont{You}} \bibnamefont{and}
  \bibinfo{author}{\bibfnamefont{F.}~\bibnamefont{Nori}},
  \bibinfo{journal}{Physics Today} \textbf{\bibinfo{volume}{58}} (11),
  \bibinfo{pages}{42} (\bibinfo{year}{2005}).

\bibitem[{\citenamefont{{Niskanen} et~al.}(2007)\citenamefont{{Niskanen},
  {Harrabi}, {Yoshihara}, {Nakamura}, {Lloyd}, and {Tsai}}}]{Niskanen2007}
\bibinfo{author}{\bibfnamefont{A.~O.} \bibnamefont{{Niskanen}}},
  \bibinfo{author}{\bibfnamefont{K.}~\bibnamefont{{Harrabi}}},
  \bibinfo{author}{\bibfnamefont{F.}~\bibnamefont{{Yoshihara}}},
  \bibinfo{author}{\bibfnamefont{Y.}~\bibnamefont{{Nakamura}}},
  \bibinfo{author}{\bibfnamefont{S.}~\bibnamefont{{Lloyd}}}, \bibnamefont{and}
  \bibinfo{author}{\bibfnamefont{J.~S.} \bibnamefont{{Tsai}}},
  \bibinfo{journal}{Science} \textbf{\bibinfo{volume}{316}},
  \bibinfo{pages}{723} (\bibinfo{year}{2007}).

\bibitem[{\citenamefont{{van der Ploeg} et~al.}(2007)\citenamefont{{van der
  Ploeg}, {Izmalkov}, {van den Brink}, {H{\"u}bner}, {Grajcar}, {Il'Ichev},
  {Meyer}, and {Zagoskin}}}]{zagoskin2007}
\bibinfo{author}{\bibfnamefont{S.~H.~W.} \bibnamefont{{van der Ploeg}}},
  \bibinfo{author}{\bibfnamefont{A.}~\bibnamefont{{Izmalkov}}},
  \bibinfo{author}{\bibfnamefont{A.~M.} \bibnamefont{{van den Brink}}},
  \bibinfo{author}{\bibfnamefont{U.}~\bibnamefont{{H{\"u}bner}}},
  \bibinfo{author}{\bibfnamefont{M.}~\bibnamefont{{Grajcar}}},
  \bibinfo{author}{\bibfnamefont{E.}~\bibnamefont{{Il'Ichev}}},
  \bibinfo{author}{\bibfnamefont{H.-G.} \bibnamefont{{Meyer}}},
  \bibnamefont{and} \bibinfo{author}{\bibfnamefont{A.~M.}
  \bibnamefont{{Zagoskin}}}, \bibinfo{journal}{Phys. Rev. Lett.}
  \textbf{\bibinfo{volume}{98}}, \bibinfo{pages}{057004}
  (\bibinfo{year}{2007}).

\bibitem[{\citenamefont{{Mariantoni} et~al.}(2008)\citenamefont{{Mariantoni},
  {Deppe}, {Marx}, {Gross}, {Wilhelm}, and {Solano}}}]{Mariantoni2008a}
\bibinfo{author}{\bibfnamefont{M.}~\bibnamefont{{Mariantoni}}},
  \bibinfo{author}{\bibfnamefont{F.}~\bibnamefont{{Deppe}}},
  \bibinfo{author}{\bibfnamefont{A.}~\bibnamefont{{Marx}}},
  \bibinfo{author}{\bibfnamefont{R.}~\bibnamefont{{Gross}}},
  \bibinfo{author}{\bibfnamefont{F.~K.} \bibnamefont{{Wilhelm}}},
  \bibnamefont{and} \bibinfo{author}{\bibfnamefont{E.}~\bibnamefont{{Solano}}},
  \bibinfo{journal}{Phys. Rev. B} \textbf{\bibinfo{volume}{78}},
  \bibinfo{pages}{104508} (\bibinfo{year}{2008}).

\bibitem[{\citenamefont{Grossmann et~al.}(1991)\citenamefont{Grossmann,
  Dittrich, Jung, and H\"anggi}}]{Grossmann1991}
\bibinfo{author}{\bibfnamefont{F.}~\bibnamefont{Grossmann}},
  \bibinfo{author}{\bibfnamefont{T.}~\bibnamefont{Dittrich}},
  \bibinfo{author}{\bibfnamefont{P.}~\bibnamefont{Jung}}, \bibnamefont{and}
  \bibinfo{author}{\bibfnamefont{P.}~\bibnamefont{H\"anggi}},
  \bibinfo{journal}{Phys. Rev. Lett.} \textbf{\bibinfo{volume}{67}},
  \bibinfo{pages}{516} (\bibinfo{year}{1991}).

\bibitem[{\citenamefont{Grossmann and H\"anggi}(1992)}]{Grossmann1992}
\bibinfo{author}{\bibfnamefont{F.}~\bibnamefont{Grossmann}} \bibnamefont{and}
  \bibinfo{author}{\bibfnamefont{P.}~\bibnamefont{H\"anggi}},
  \bibinfo{journal}{Europhys. Lett.} \textbf{\bibinfo{volume}{18}},
  \bibinfo{pages}{571} (\bibinfo{year}{1992}).

\bibitem[{\citenamefont{Grifoni and H\"anggi}(1998)}]{GrifoniPR}
\bibinfo{author}{\bibfnamefont{M.}~\bibnamefont{Grifoni}} \bibnamefont{and}
  \bibinfo{author}{\bibfnamefont{P.}~\bibnamefont{H\"anggi}},
  \bibinfo{journal}{Phys. Rep.} \textbf{\bibinfo{volume}{304}},
  \bibinfo{pages}{229} (\bibinfo{year}{1998}).

\bibitem[{\citenamefont{Platero and Aguado}(2004)}]{PlateroPR}
\bibinfo{author}{\bibfnamefont{G.}~\bibnamefont{Platero}} \bibnamefont{and}
  \bibinfo{author}{\bibfnamefont{R.}~\bibnamefont{Aguado}},
  \bibinfo{journal}{Phys. Rep.} \textbf{\bibinfo{volume}{395}},
  \bibinfo{pages}{1} (\bibinfo{year}{2004}).

\bibitem[{\citenamefont{Kenkre and Raghavan}(2000)}]{kenkre2000}
\bibinfo{author}{\bibfnamefont{V.~M.} \bibnamefont{Kenkre}} \bibnamefont{and}
  \bibinfo{author}{\bibfnamefont{S.}~\bibnamefont{Raghavan}},
  \bibinfo{journal}{J. Opt. B: Quantum Semiclass. Opt.}
  \textbf{\bibinfo{volume}{2}}, \bibinfo{pages}{686} (\bibinfo{year}{2000}).

\bibitem[{\citenamefont{Longhi}(2008)}]{Longhi2008}
\bibinfo{author}{\bibfnamefont{S.}~\bibnamefont{Longhi}},
  \bibinfo{journal}{Phys. Rev. B} \textbf{\bibinfo{volume}{77}},
  \bibinfo{pages}{195326} (\bibinfo{year}{2008}).

\bibitem[{\citenamefont{Kohler et~al.}(2005)\citenamefont{Kohler, Lehmann, and
  H\"anggi}}]{Kohler2005a}
\bibinfo{author}{\bibfnamefont{S.}~\bibnamefont{Kohler}},
  \bibinfo{author}{\bibfnamefont{J.}~\bibnamefont{Lehmann}}, \bibnamefont{and}
  \bibinfo{author}{\bibfnamefont{P.}~\bibnamefont{H\"anggi}},
  \bibinfo{journal}{Phys. Rep.} \textbf{\bibinfo{volume}{406}},
  \bibinfo{pages}{379} (\bibinfo{year}{2005}).

\bibitem[{\citenamefont{Kayanuma and Saito}(2008)}]{Kayanuma2008}
\bibinfo{author}{\bibfnamefont{Y.}~\bibnamefont{Kayanuma}} \bibnamefont{and}
  \bibinfo{author}{\bibfnamefont{K.}~\bibnamefont{Saito}},
  \bibinfo{journal}{Phys. Rev. A} \textbf{\bibinfo{volume}{77}},
  \bibinfo{pages}{010101(R)} (\bibinfo{year}{2008}).

\bibitem[{\citenamefont{Camalet et~al.}(2003)\citenamefont{Camalet, Lehmann,
  Kohler, and H\"anggi}}]{Camalet2003a}
\bibinfo{author}{\bibfnamefont{S.}~\bibnamefont{Camalet}},
  \bibinfo{author}{\bibfnamefont{J.}~\bibnamefont{Lehmann}},
  \bibinfo{author}{\bibfnamefont{S.}~\bibnamefont{Kohler}}, \bibnamefont{and}
  \bibinfo{author}{\bibfnamefont{P.}~\bibnamefont{H\"anggi}},
  \bibinfo{journal}{Phys. Rev. Lett.} \textbf{\bibinfo{volume}{90}},
  \bibinfo{pages}{210602} (\bibinfo{year}{2003}).

\bibitem[{\citenamefont{{Creffield}}(2007)}]{Creffield2007}
\bibinfo{author}{\bibfnamefont{C.~E.} \bibnamefont{{Creffield}}},
  \bibinfo{journal}{Phys. Rev. Lett.} \textbf{\bibinfo{volume}{99}},
  \bibinfo{pages}{110501} (\bibinfo{year}{2007}).

\bibitem[{\citenamefont{Burgarth}(2006)}]{Burgarth2006}
\bibinfo{author}{\bibfnamefont{D.}~\bibnamefont{Burgarth}}, Ph.D. thesis,
  \bibinfo{school}{University College London} (\bibinfo{year}{2006}),
  \bibinfo{note}{arXiv:0704.1309 [quant-ph]}.

\bibitem[{\citenamefont{Ciaramicoli et~al.}(2003)\citenamefont{Ciaramicoli,
  Marzoli, and Tombesi}}]{Ciaramicoli2003}
\bibinfo{author}{\bibfnamefont{G.}~\bibnamefont{Ciaramicoli}},
  \bibinfo{author}{\bibfnamefont{I.}~\bibnamefont{Marzoli}}, \bibnamefont{and}
  \bibinfo{author}{\bibfnamefont{P.}~\bibnamefont{Tombesi}},
  \bibinfo{journal}{Phys. Rev. Lett.} \textbf{\bibinfo{volume}{91}},
  \bibinfo{pages}{017901} (\bibinfo{year}{2003}).

\bibitem[{\citenamefont{Stahl et~al.}(2005)\citenamefont{Stahl, Alonso, Djekic,
  Quint, Valenzuela, Verdu, Vogel, and Werth}}]{Stahl2005}
\bibinfo{author}{\bibfnamefont{S.}~\bibnamefont{Stahl}},
  \bibinfo{author}{\bibfnamefont{F.~G.~J.} \bibnamefont{Alonso}},
  \bibinfo{author}{\bibfnamefont{S.}~\bibnamefont{Djekic}},
  \bibinfo{author}{\bibfnamefont{W.}~\bibnamefont{Quint}},
  \bibinfo{author}{\bibfnamefont{T.}~\bibnamefont{Valenzuela}},
  \bibinfo{author}{\bibfnamefont{J.}~\bibnamefont{Verdu}},
  \bibinfo{author}{\bibfnamefont{M.}~\bibnamefont{Vogel}}, \bibnamefont{and}
  \bibinfo{author}{\bibfnamefont{G.}~\bibnamefont{Werth}},
  \bibinfo{journal}{Eur. Phys. J. D} \textbf{\bibinfo{volume}{32}},
  \bibinfo{pages}{139} (\bibinfo{year}{2005}).

\bibitem[{\citenamefont{{Helmer} et~al.}(2009)\citenamefont{{Helmer},
  {Mariantoni}, {Fowler}, {von Delft}, {Solano}, and
  {Marquardt}}}]{Helmer2007a}
\bibinfo{author}{\bibfnamefont{F.}~\bibnamefont{{Helmer}}},
  \bibinfo{author}{\bibfnamefont{M.}~\bibnamefont{{Mariantoni}}},
  \bibinfo{author}{\bibfnamefont{A.~G.} \bibnamefont{{Fowler}}},
  \bibinfo{author}{\bibfnamefont{J.}~\bibnamefont{{von Delft}}},
  \bibinfo{author}{\bibfnamefont{E.}~\bibnamefont{{Solano}}}, \bibnamefont{and}
  \bibinfo{author}{\bibfnamefont{F.}~\bibnamefont{{Marquardt}}},
  \bibinfo{journal}{EPL} \textbf{\bibinfo{volume}{85}}, \bibinfo{pages}{50007}
  (\bibinfo{year}{2009}).

\bibitem[{\citenamefont{Valle et~al.}(2007)\citenamefont{Valle, Ornigotti,
  Cianci, Foglietti, Laporta, and Longhi}}]{Longhi2007}
\bibinfo{author}{\bibfnamefont{G.~D.} \bibnamefont{Valle}},
  \bibinfo{author}{\bibfnamefont{M.}~\bibnamefont{Ornigotti}},
  \bibinfo{author}{\bibfnamefont{E.}~\bibnamefont{Cianci}},
  \bibinfo{author}{\bibfnamefont{V.}~\bibnamefont{Foglietti}},
  \bibinfo{author}{\bibfnamefont{P.}~\bibnamefont{Laporta}}, \bibnamefont{and}
  \bibinfo{author}{\bibfnamefont{S.}~\bibnamefont{Longhi}},
  \bibinfo{journal}{Phys. Rev. Lett.} \textbf{\bibinfo{volume}{98}},
  \bibinfo{pages}{263601} (\bibinfo{year}{2007}).

\bibitem[{\citenamefont{Bennett et~al.}(1993)\citenamefont{Bennett, Brassard,
  Crepeau, Jozsa, Peres, and Wootters}}]{teleport}
\bibinfo{author}{\bibfnamefont{C.~H.} \bibnamefont{Bennett}},
  \bibinfo{author}{\bibfnamefont{G.}~\bibnamefont{Brassard}},
  \bibinfo{author}{\bibfnamefont{C.}~\bibnamefont{Crepeau}},
  \bibinfo{author}{\bibfnamefont{R.}~\bibnamefont{Jozsa}},
  \bibinfo{author}{\bibfnamefont{A.}~\bibnamefont{Peres}}, \bibnamefont{and}
  \bibinfo{author}{\bibfnamefont{W.}~\bibnamefont{Wootters}},
  \bibinfo{journal}{Phys. Rev. Lett.} \textbf{\bibinfo{volume}{70}},
  \bibinfo{pages}{1895} (\bibinfo{year}{1993}).

\bibitem[{\citenamefont{Galve et~al.}(2009)\citenamefont{Galve, Zueco, Kohler,
  Lutz, and H\"anggi}}]{Galve2009a}
\bibinfo{author}{\bibfnamefont{F.}~\bibnamefont{Galve}},
  \bibinfo{author}{\bibfnamefont{D.}~\bibnamefont{Zueco}},
  \bibinfo{author}{\bibfnamefont{S.}~\bibnamefont{Kohler}},
  \bibinfo{author}{\bibfnamefont{E.}~\bibnamefont{Lutz}}, \bibnamefont{and}
  \bibinfo{author}{\bibfnamefont{P.}~\bibnamefont{H\"anggi}},
  \bibinfo{journal}{Phys. Rev. A} \textbf{\bibinfo{volume}{79}}
  (\bibinfo{year}{2009}).

\bibitem[{\citenamefont{{Wichterich} and {Bose}}(2008)}]{bose}
\bibinfo{author}{\bibfnamefont{H.}~\bibnamefont{{Wichterich}}}
  \bibnamefont{and} \bibinfo{author}{\bibfnamefont{S.}~\bibnamefont{{Bose}}}
  (\bibinfo{year}{2008}), \eprint{arXiv:0806.4568 [quant-ph]}.

\bibitem[{\citenamefont{Astumian and H\"anggi}(2002)}]{Astumian2002}
\bibinfo{author}{\bibfnamefont{R.~D.} \bibnamefont{Astumian}} \bibnamefont{and}
  \bibinfo{author}{\bibfnamefont{P.}~\bibnamefont{H\"anggi}},
  \bibinfo{journal}{Physics Today} \textbf{\bibinfo{volume}{55}} (11),
  \bibinfo{pages}{33} (\bibinfo{year}{2002}).

\bibitem[{\citenamefont{H\"anggi and Marchesoni}(2009)}]{RMP2009}
\bibinfo{author}{\bibfnamefont{P.}~\bibnamefont{H\"anggi}} \bibnamefont{and}
  \bibinfo{author}{\bibfnamefont{F.}~\bibnamefont{Marchesoni}},
  \bibinfo{journal}{Rev. Mod. Phys.} \textbf{\bibinfo{volume}{81}},
  \bibinfo{pages}{387} (\bibinfo{year}{2009}).

\bibitem[{\citenamefont{{Wootters}}(1998)}]{Wootters1998}
\bibinfo{author}{\bibfnamefont{W.~K.} \bibnamefont{{Wootters}}},
  \bibinfo{journal}{Phys. Rev. Lett.} \textbf{\bibinfo{volume}{80}},
  \bibinfo{pages}{2245} (\bibinfo{year}{1998}).

\bibitem[{\citenamefont{Roos et~al.}(2000)\citenamefont{Roos, Leibfried, Mundt,
  Schmidt-Kaler, Eschner, and Blatt}}]{Roos2000a}
\bibinfo{author}{\bibfnamefont{C.~F.} \bibnamefont{Roos}},
  \bibinfo{author}{\bibfnamefont{D.}~\bibnamefont{Leibfried}},
  \bibinfo{author}{\bibfnamefont{A.}~\bibnamefont{Mundt}},
  \bibinfo{author}{\bibfnamefont{F.}~\bibnamefont{Schmidt-Kaler}},
  \bibinfo{author}{\bibfnamefont{J.}~\bibnamefont{Eschner}}, \bibnamefont{and}
  \bibinfo{author}{\bibfnamefont{R.}~\bibnamefont{Blatt}},
  \bibinfo{journal}{Phys. Rev. Lett.} \textbf{\bibinfo{volume}{85}},
  \bibinfo{pages}{5547} (\bibinfo{year}{2000}).

\bibitem[{\citenamefont{Leibfried et~al.}(2003)\citenamefont{Leibfried, Blatt,
  Monroe, and Wineland}}]{wineland}
\bibinfo{author}{\bibfnamefont{D.}~\bibnamefont{Leibfried}},
  \bibinfo{author}{\bibfnamefont{R.}~\bibnamefont{Blatt}},
  \bibinfo{author}{\bibfnamefont{C.}~\bibnamefont{Monroe}}, \bibnamefont{and}
  \bibinfo{author}{\bibfnamefont{D.}~\bibnamefont{Wineland}},
  \bibinfo{journal}{Rev. Mod. Phys.} \textbf{\bibinfo{volume}{75}},
  \bibinfo{pages}{281} (\bibinfo{year}{2003}).

\bibitem[{\citenamefont{Magalinski\u{\i}}(1959)}]{magalinskii1959}
\bibinfo{author}{\bibfnamefont{V.~B.} \bibnamefont{Magalinski\u{\i}}},
  \bibinfo{journal}{Sov. Phys. JETP} \textbf{\bibinfo{volume}{9}},
  \bibinfo{pages}{1381} (\bibinfo{year}{1959}).

\bibitem[{\citenamefont{Caldeira and Leggett}(1983)}]{Caldeira1983}
\bibinfo{author}{\bibfnamefont{A.~O.} \bibnamefont{Caldeira}} \bibnamefont{and}
  \bibinfo{author}{\bibfnamefont{A.~L.} \bibnamefont{Leggett}},
  \bibinfo{journal}{Ann. Phys. (N.Y.)} \textbf{\bibinfo{volume}{149}},
  \bibinfo{pages}{374} (\bibinfo{year}{1983}).

\bibitem[{\citenamefont{H\"anggi and Ingold}(2005)}]{Chaos2005}
\bibinfo{author}{\bibfnamefont{P.}~\bibnamefont{H\"anggi}} \bibnamefont{and}
  \bibinfo{author}{\bibfnamefont{G.-L.} \bibnamefont{Ingold}},
  \bibinfo{journal}{Chaos} \textbf{\bibinfo{volume}{15}},
  \bibinfo{pages}{026105} (\bibinfo{year}{2005}).

\bibitem[{\citenamefont{Blum}(1996)}]{Blum1996a}
\bibinfo{author}{\bibfnamefont{K.}~\bibnamefont{Blum}},
  \emph{\bibinfo{title}{Density Matrix Theory and Applications}}
  (\bibinfo{publisher}{Springer}, \bibinfo{address}{New York},
  \bibinfo{year}{1996}), \bibinfo{edition}{2nd} ed.

\bibitem[{\citenamefont{Lindblad}(1976)}]{Lindblad1976a}
\bibinfo{author}{\bibfnamefont{G.}~\bibnamefont{Lindblad}},
  \bibinfo{journal}{Commun. Math. Phys.} \textbf{\bibinfo{volume}{48}},
  \bibinfo{pages}{119} (\bibinfo{year}{1976}).

\bibitem[{\citenamefont{{Doll} et~al.}(2008)\citenamefont{{Doll}, {Zueco},
  {Wubs}, {Kohler}, and {H{\"a}nggi}}}]{Doll2008}
\bibinfo{author}{\bibfnamefont{R.}~\bibnamefont{{Doll}}},
  \bibinfo{author}{\bibfnamefont{D.}~\bibnamefont{{Zueco}}},
  \bibinfo{author}{\bibfnamefont{M.}~\bibnamefont{{Wubs}}},
  \bibinfo{author}{\bibfnamefont{S.}~\bibnamefont{{Kohler}}}, \bibnamefont{and}
  \bibinfo{author}{\bibfnamefont{P.}~\bibnamefont{{H{\"a}nggi}}},
  \bibinfo{journal}{Chemical Physics} \textbf{\bibinfo{volume}{347}},
  \bibinfo{pages}{243} (\bibinfo{year}{2008}).

\bibitem[{\citenamefont{{Roos} et~al.}(1999)\citenamefont{{Roos}, {Zeiger},
  {Rohde}, {N{\"a}gerl}, {Eschner}, {Leibfried}, {Schmidt-Kaler}, and
  {Blatt}}}]{Roos1999}
\bibinfo{author}{\bibfnamefont{C.}~\bibnamefont{{Roos}}},
  \bibinfo{author}{\bibfnamefont{T.}~\bibnamefont{{Zeiger}}},
  \bibinfo{author}{\bibfnamefont{H.}~\bibnamefont{{Rohde}}},
  \bibinfo{author}{\bibfnamefont{H.~C.} \bibnamefont{{N{\"a}gerl}}},
  \bibinfo{author}{\bibfnamefont{J.}~\bibnamefont{{Eschner}}},
  \bibinfo{author}{\bibfnamefont{D.}~\bibnamefont{{Leibfried}}},
  \bibinfo{author}{\bibfnamefont{F.}~\bibnamefont{{Schmidt-Kaler}}},
  \bibnamefont{and} \bibinfo{author}{\bibfnamefont{R.}~\bibnamefont{{Blatt}}},
  \bibinfo{journal}{Phys. Rev. Lett.} \textbf{\bibinfo{volume}{83}},
  \bibinfo{pages}{4713} (\bibinfo{year}{1999}).

\bibitem[{\citenamefont{Doll et~al.}(2007)\citenamefont{Doll, Wubs, H\"anggi,
  and Kohler}}]{Doll2007a}
\bibinfo{author}{\bibfnamefont{R.}~\bibnamefont{Doll}},
  \bibinfo{author}{\bibfnamefont{M.}~\bibnamefont{Wubs}},
  \bibinfo{author}{\bibfnamefont{P.}~\bibnamefont{H\"anggi}}, \bibnamefont{and}
  \bibinfo{author}{\bibfnamefont{S.}~\bibnamefont{Kohler}},
  \bibinfo{journal}{Phys. Rev. B} \textbf{\bibinfo{volume}{76}},
  \bibinfo{pages}{045317} (\bibinfo{year}{2007}).

\end{thebibliography}

\end{document}